\documentclass{aa}  
\usepackage{graphicx}
\usepackage[compatibility=false]{caption}
\usepackage{txfonts}
\usepackage{booktabs}
\usepackage{siunitx}
\usepackage{color}
\usepackage[dvipsnames]{xcolor}
\usepackage{gensymb}
\usepackage{natbib,twoopt}
\usepackage{lscape}
\usepackage{gensymb}
\usepackage{caption}
\usepackage{subcaption}
\usepackage[symbol]{footmisc} 
\usepackage{tablefootnote}
\usepackage{soul}
\usepackage{comment}
\usepackage[bookmarks=true, pdfnewwindow=true, colorlinks=true, allcolors=blue, breaklinks=true, final=true]{hyperref}
\usepackage{orcidlink}
\bibpunct{(}{)}{;}{a}{}{,} 

\def\purple#1 {{\textcolor{purple}{#1}}\ }
\def\red#1 {\textcolor{red}{#1}}
\def\blue#1 {{\textcolor{blue}{#1}}\ }
\def\zy#1 {\textcolor{forestgreen}{ zy: #1}}

\def\reference#1 {\textcolor{red}{ reference: #1}}

\begin{document}

\title{Constraining Dust Properties Along the Minor Axis of Andromeda}

   \author{Sihan Jiao
          \inst{1,2}
          \and
          Linjing Feng\inst{2,3}
          \and
          Henrik Beuther\inst{1}
          \and
          Hauyu Baobab Liu\inst{4,5}
          \and
          Yuxin Lin\inst{6}
          \and
          Jingwen Wu\inst{3,2}
          \and
          Chao-Wei Tsai\inst{2,7,3}
          \and
          Di Li\inst{8,2}
          \and
          Kiyoaki Doi\inst{1}
          \and
          Jakob den Brok\inst{1}
          \and
          Fengwei Xu\inst{1}
          \and
          Sim\'{o}n D. Reyes-Reyes\inst{1}
          \and
          Carsten Kramer\inst{9}
          \and
          Ryan P. Keenan\inst{1}
          \and
          Xiangyu Zhang\inst{1}
          \and
          Fanyi Meng\inst{8}
          \and
          Zongnan Li\inst{10,11}
          \and
          Fangyuan Deng\inst{3,2}
          \and
          Yuxiang Liu\inst{3,2}
}

    \institute{Max Planck Institute for Astronomy, Konigstuhl 17, D-69117 Heidelberg, Germany\\
    \email{sijiao@mpia.de}
            \and
            National Astronomical Observatories, Chinese Academy of Sciences, Beijing 100101, China\\
        \email{sihanjiao@nao.cas.cn; 	ljfeng@nao.cas.cn}
            \and
            University of Chinese Academy of Sciences, Beijing 100049, China 
            \and
            Department of Physics, National Sun Yat-Sen University, No. 70, Lien-Hai Road, Kaohsiung City 80424, Taiwan, R.O.C. 
            \and
            Center of Astronomy and Gravitation, National Taiwan Normal University, Taipei 116, Taiwan 
            \and
            Max-Planck-Institut f\"ur Extraterrestrische Physik, Giessenbachstr. 1, D-85748 Garching bei M\"unchen, Germany
            \and
            Institute for Frontiers in Astronomy and Astrophysics, Beijing Normal University,  Beijing 102206, China
            \and
            New Cornerstone Science Laboratory, Department of Astronomy, Tsinghua University, Beijing 100084, China
            \and
            Institut de RadioAstronomie Millim\'etrique (IRAM), 38406 Saint-Martin-d'H\`eres, France
            \and
            Korea Astronomy and Space Science Institute, 776 Daedeok-daero, Yuseong-gu, Daejeon 34055, Republic of Korea
            \and
            East Asian Core Observatories Association (EACOA) Fellow
}

\abstract
{Previous studies of dust emission in M31 based primarily on {\it Herschel} data have reported inconsistent radial trends in the dust emissivity index ($\beta$), partly due to limited wavelength coverage and the intrinsic $T_{\rm dust}$–$\beta$ degeneracy.}
{We aim to improve the constraints of $\beta$ determinations and obtain the spatial surface density and temperature distributions at GMC scales in M31 by extending the spectral coverage into the millimeter regime and by quantifying the impact of wavelength coverage on the $T_{\rm dust}$–$\beta$ degeneracy.}
{We present new IRAM 30m NIKA2 1.15 mm observations along the minor axis of M31 at an angular resolution of $11''$, corresponding to $\sim$40 pc at the distance of M31. 
These data are used with {\it Herschel} and JCMT SCUBA-2 maps to perform pixel-by-pixel single-component modified blackbody fitting. 
}
{Including the 850 $\mu$m and 1.15 mm data helps reduce the $T_{\rm dust}$–$\beta$ degeneracy and yields an average emissivity index of $\beta \approx 1.8$, systematically lower than values derived from far-infrared–only analyses. 
Radial analysis shows higher $\beta$ values in the 5 kpc ring compared to the 10 kpc ring.
The elevated $\beta$ values in the inner 5 kpc may be linked to changes in dust composition, potentially associated with enhanced starlight heating in this region.}
{Millimeter wavelength coverage is important for improving $\beta$ constraints in M31. 
The lower $\beta$ values inferred from the 250--1150 $\mu$m SED fitting suggest possible variations in dust emissivity and highlight the importance of sampling the long-wavelength SED slope in nearby galaxies.
}


\maketitle
\titlerunning{M31 NIKA2 Pilot Map}
\authorrunning{Jiao et al.}

%

\section{Introduction}

Dust constitutes only a small fraction ($\sim$1\%) of the interstellar medium
(ISM), yet it plays a crucial role in various processes within galaxies \citep{Draine2011}, including heating the gas through the photoelectric effect \citep{Draine1978ApJS}, cooling gas via infrared emission, providing surfaces for molecular gas formation, and shielding molecules from dissociating radiation \citep{Gould1963ApJ,Hollenbach1971ApJ}.
Understanding the dust properties within galaxies is therefore essential for studying the life cycle of gas in the ISM, complementing studies of the multiphase gas based on tracers such as H\,{\sc i}, CO, and [C\,{\sc i}] \citep[e.g.,][]{Li&Goldsmith2003ApJ,Bolatto2013,Li2018ApJS,Yu2024SCPMA,Xia2026SCPMA}.

Our understanding of dust properties has primarily been shaped by infrared observations in the Milky Way \citep[e.g.,][]{Molinari2010,Planck2014A&A...571A..11P,Csengeri2016,Planck2016A&A...586A.132P}, which reveal that dust properties may vary in different ISM environments.
However, the connection between these variations to large-scale galactic structures, such as spiral arms and rings, remains poorly understood. 
The edge-on projection of the Milky Way and the ambiguity of line-of-sight distances pose significant challenges to achieving a comprehensive, panoramic view of dust properties. 

Studies of nearby galaxies can overcome these challenges \citep[e.g.,][]{Galliano2018,Tabatabaei2014,Ejlali2025,Pantoni2026}, avoiding the foreground/background confusion and distance ambiguities inherent to Galactic observations.
The Andromeda galaxy (M31), located at a distance of 766$\pm$18 kpc \citep[][]{Lee2023ApJ}, is the nearest massive spiral galaxy.
Its proximity provides an opportunity to study the dust properties in an external spiral galaxy at high spatial resolution with millimeter single-dish telescopes.
Extensive infrared and submillimeter observations of M31 have been obtained with {\it IRAS} \citep{Habing1984ApJ,Devereux1994AJ}, the {\it Infrared Space Observatory} \citep{Haas1998A&A}, the {\it Spitzer Space Telescope} \citep{Gordon2006ApJ}, {\it Herschel Space Observatory} \citep{Fritz2012A&A,Groves2012MNRAS,Draine2014ApJ}, {\it Planck} space telescope \citep{Planck2015A&A}, and JCMT-SCUBA2 \citep{SmithApJS,Jiao2025am31}.
Based on these observations, the dust properties across the disk have been studied in detail \citep[e.g.,][]{Smith2012ApJ,Draine2014ApJ,Planck2015A&A,Jiao2025am31}.
In many of these analyses, the dust emission is modeled using a single modified blackbody, $S_{\nu}=\Omega B_{\nu}(T_{d})(1-e^{-\tau_{\nu}})$, where $S_{\nu}$ is the observed flux, $\Omega$ is the solid angle of the emission region, $B_{\nu}(T_{d})$ is the Planck function at temperature $T_{d}$, and $\tau_{\nu}$ is the optical depth of dust.
The optical depth is a product of the dust mass opacity ($\kappa_{\nu}$) and dust surface density ($\Sigma_{\mbox{\scriptsize dust}}$).
Since dust grains cannot emit or absorb efficiently at wavelengths that are much longer than their size, the dust mass opacity $\kappa_{\nu}$ is approximately $\kappa_{\nu}=\kappa_{\mbox{\tiny{857 GHz}}}(\nu/857\,\mbox{[GHz]})^{\beta}$, where $\beta$ is known as the dust opacity spectral index.
In the diffuse ISM of the solar neighborhood, $\beta \sim 2$, while grain growth processes, such as those observed in protoplanetary disks, can lead to lower $\beta$ values \citep[e.g.,][]{Draine2006ApJ,Carrasco2019ApJ,Liu2021ApJ,jiang2024A&A...682A..32J}.
Under the assumption of an isothermal ISM, three well-sampled wavelength bands are, in principle, sufficient to constrain the three free parameters: $T_{d}$, $\Sigma$, and $\beta$. 
However, dust within the telescope beam and along the line of sight generally spans a range of temperatures.
The greater contribution of warmer dust at shorter wavelengths broadens the observed SED, and fitting such emission with a single modified blackbody may bias the derived dust temperature toward higher values and the emissivity index $\beta$ toward lower values \citep[e.g.,][]{Shetty2009ApJb,JuvelaYsard2012,Galliano2018}.

\begin{figure*}[ht!]
\hspace{-0.8cm}
\includegraphics[width=19.9cm]{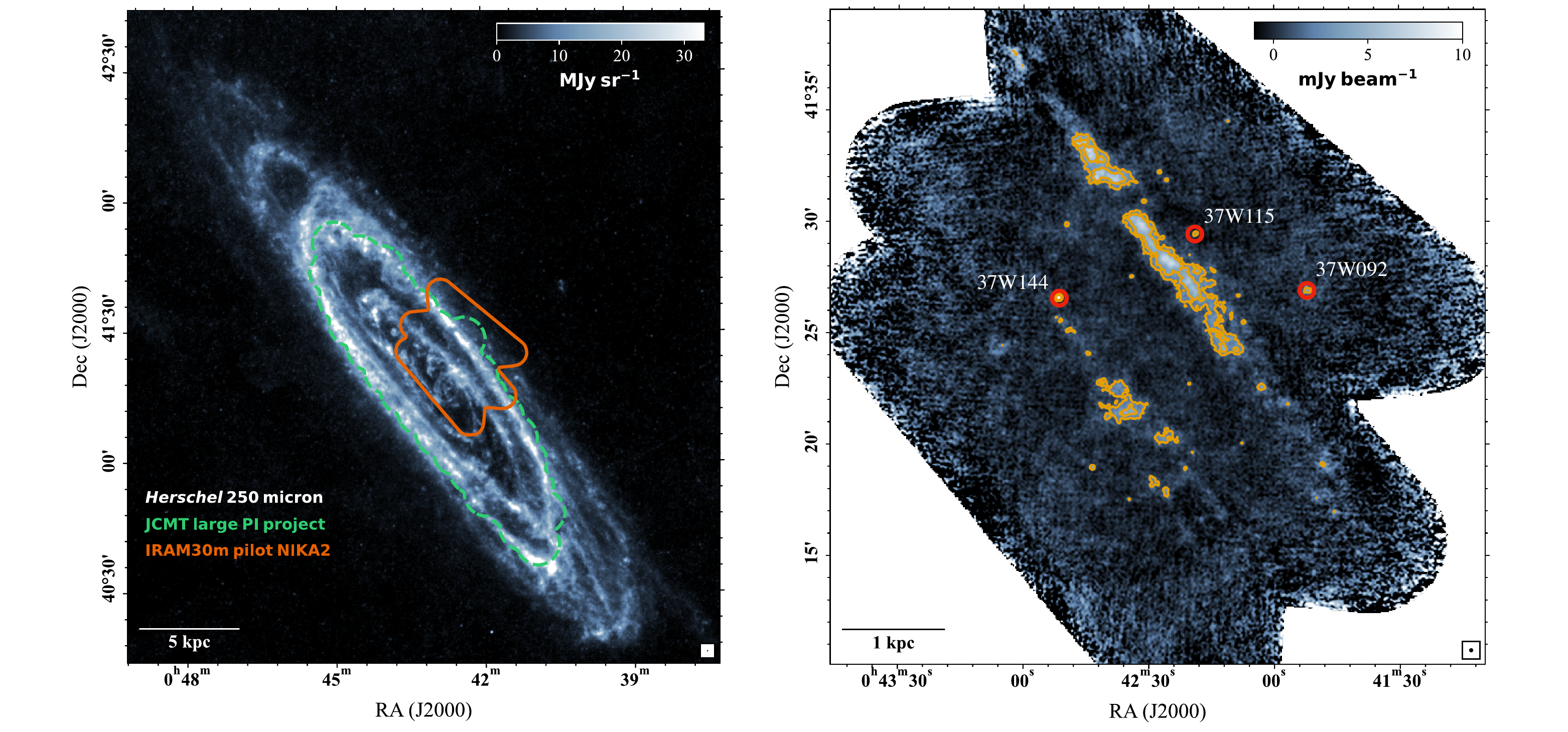}
\caption{
{\it Left:--} Overview of the datasets used in this work.
The background color map shows the \textit{Herschel} 250 $\mu$m image \citep{Fritz2012A&A,Smith2012ApJ}. 
The footprints of the JCMT large PI project \citep{Jiao2025am31} and the IRAM 30m NIKA2 pilot project are shown with green dashed and orange lines, respectively.
{\it Right:--}The new M31 1.15 mm pilot NIKA2 map, with an angular resolution of 11.$''$6.
Orange contours denote regions with signal-to-noise ratios of 5 and 10, derived from the flux and noise maps.
Red circles mark masked strong non-thermal radio sources.
}
\label{fig:data}
\end{figure*}

However, the reported dust property distributions of M31 in previous works are mixed, especially regarding radial trends in $\beta$ \citep[e.g.,][]{Smith2012ApJ,Draine2014ApJ,Planck2015A&A,Whitworth2019MNRAS,Athikkat-Eknath2022MNRAS,Jiao2025am31}.
Based on {\it Herschel} observations, \cite{Smith2012ApJ} identified a sharp change in $\beta$ at R$\approx$3 kpc, with $\beta$ rising from 2 to 2.5 within 3\,kpc and then decreasing from 2.5 to 1.5 from $\sim$3\,kpc to the $\sim$ 15 kpc radii.
\citet{Mattsson2014MNRAS} proposed that substantial interstellar dust grain growth might account for this apparent drop in $\beta$.
In contrast, \citet{Draine2014ApJ} derived $\beta$ from {\it Herschel} SPIRE500/SPIRE250 band ratios and found $\beta \approx 2.08$ for $R \gtrsim 7$ kpc, consistent with the dust model in \cite{Draine2007ApJ}.
Including JCMT-SCUBA2 450 and 850 $\mu$m observations in the SED fitting, \cite{Jiao2025am31} reported smaller variations in $\beta$ (1.5 $\lesssim$ $\beta$ $\lesssim$ 2.3), especially for R$\lesssim$7\,kpc. 
The results show an average $\beta$ value of 1.8, which is smaller than values reported in other studies, such as, \cite{Draine2014ApJ}, where $\beta$ was defined in a post-processing step, between 250 and 500 $\mu$m, and \cite{Whitworth2019MNRAS}, where $\beta$ was treated as an intrinsic parameter in the \textsc{PPMAP} \citep{Marsh2015MNRAS} analysis.
These discrepancies may arise in the differences in wavelength coverage and fitting methodology.

In the modified blackbody SED fitting, when signal-to-noise ratio (S/N) is limited, the observations at $<$0.5 mm wavelengths provide better constraints on $T_{d}$; at longer wavelength bands, $B_{\nu}(T_{d})$ has an approximated simple $\nu^{2+\beta}$ dependence (i.e., Rayleigh-Jeans limit) which makes the measurement of $\beta$ more robust.
However, most previous studies of M31 rely mainly on {\it Herschel} data, which do not sample the Rayleigh–Jeans tail well for dust at typical temperatures of $\sim$18 K. 
The absence of longer-wavelength coverage may therefore contribute to systematic uncertainties in earlier dust property analyses.

To address this limitation and directly test how dust properties depend on large-scale galactic structures, from the high-temperature central regions to the colder outer disk, we conducted pilot IRAM 30m NIKA2 \citep{Adam2018A&A,Perotto2020A&A} observations covering the disk out to a radius of $\sim$15 kpc along the minor axis of M31. 
By comparing with the JCMT-SCUBA2, {\it Planck}, and {\it Herschel} observations, we aim to carry out a state-of-the-art study on the dust property and their environmental dependence in M31 at the giant molecular cloud (GMC) scale.
As a pilot, this project will provide immediate scientific results and establish a framework for future large field of view NIKA2 survey of molecular clouds across the M31 disk.

This paper is organized as follows. 
Section \ref{sec:obs} outlines the observations and the methodologies employed to analyze the data.
Section \ref{sec:res} presents a detailed analysis of the dust properties of the pilot region.
The discussion on the derived dust properties is introduced in Section \ref{sec:dis}.
Finally, our main findings are summarized in Section \ref{sec:con}.

\section{Observations and Methods}\label{sec:obs}

\begin{figure*}[ht!]
\vspace{-0.3cm}
\hspace{0.5cm}
\includegraphics[width=16.7cm,]{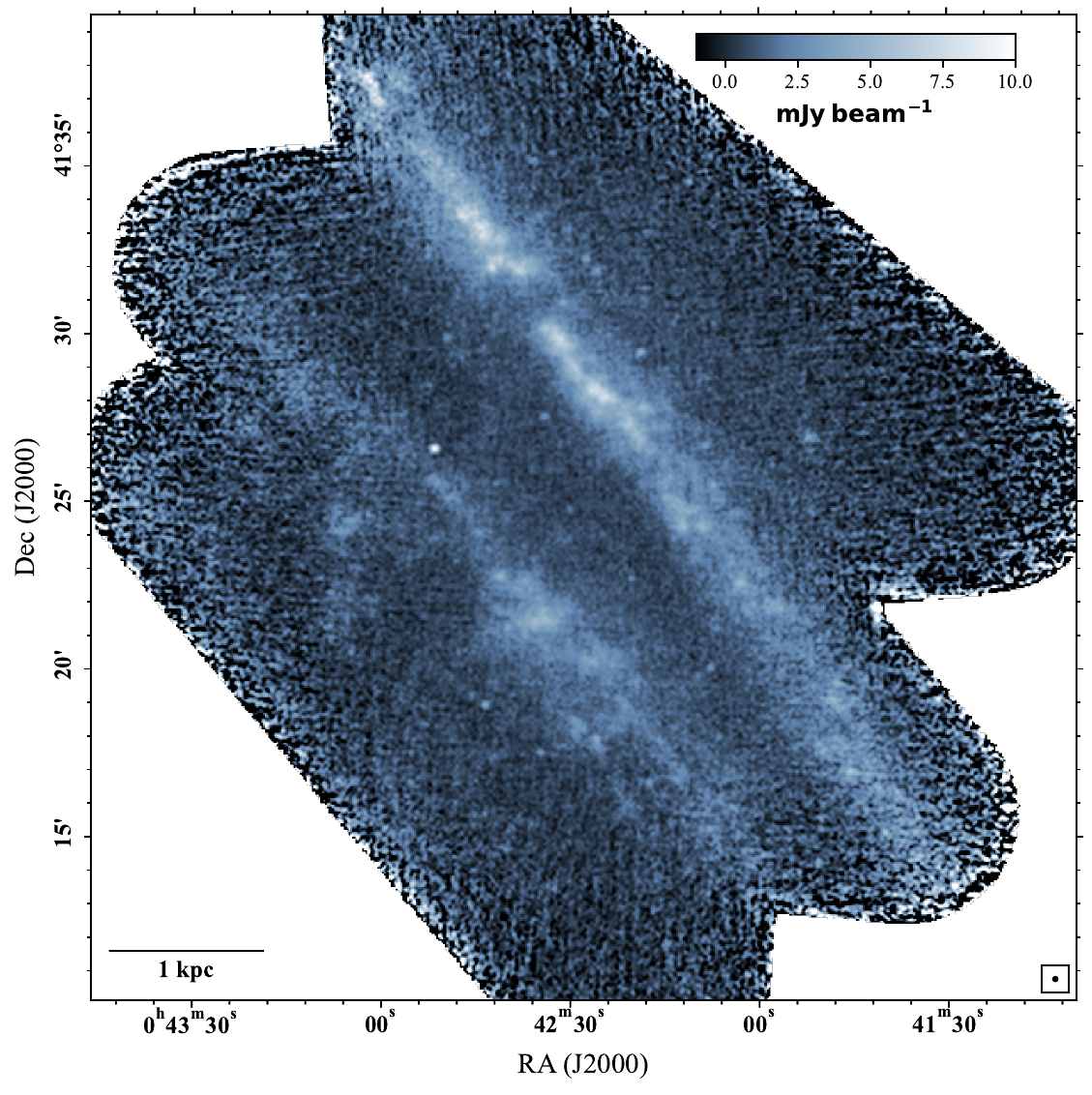}
\caption{
The 11.$''$6 resolution, 1.15 mm continuum image of M31 produced by combining \textit{Planck} and NIKA2 observations.
}
\label{fig:nika2_1mm}
\end{figure*}

\subsection{IRAM 30m Telescope NIKA2 Observations}\label{subsec:obs_nika2}

To constrain dust properties along the minor axis of M31, we used two 14$'$ $\times$ 12$'$ total power mode OTF maps (right panel of Figure \ref{fig:data}) with a scanning speed of 60$''$ s$^{-1}$.
The observations were carried out with the New IRAM KIDs Array 2 \citep[NIKA2,][]{Adam2018A&A,Perotto2020A&A} on the IRAM 30m telescope in November 2025 (project 066-25, PI: Sihan Jiao), covering a wide sector from the inner 5 kpc ring to the outer 10 kpc ring.
NIKA2 consists of 2900 kinetic inductance detectors (KIDs) operating simultaneously at 150 and 260 GHz. 
Given that the primary goal of this study is to constrain thermal dust emission at GMC scales, we focus exclusively on the 260 GHz (1.15 mm) observations, which provide a spatial resolution of $\sim$40 pc at the distance of M31 \citep[766$\pm$18 kpc,][]{Lee2023ApJ}.
The 150 GHz data will be analyzed in a future study.

The NIKA2 data were reduced using the PIIC software package\footnote{PIIC handbook: \url{https://www.iram.fr/~gildas/dist/piic.pdf}.}.
Before projection onto the map grid, the detector time streams were calibrated and corrected for atmospheric extinction, and processed to remove correlated sky noise and residual instrumental drifts.
The 260 GHz data from both OTF maps were processed jointly using an iterative data-reduction procedure to remove the time-variable atmospheric emission.
In each iteration, the signal-to-noise ratio per pixel is calculated, and pixels close to a predetermined threshold are identified. 
This creates a source map that is subtracted from the data timeline in the next cycle of the iteration loop, allowing for improved source definition.
The iterative reduction was stopped after 20 iterations to obtain the final 1.15 mm map, when the peak flux density difference between the previous two iterations fell below 0.5\%. 
The final map is centered at (RA, Dec) = (00:42:27.9, +41:24:48.4) and is sampled on a $735 \times 735$ pixel grid with a pixel scale of $3''$.
The map has an angular resolution of $11.6''$ and achieves a typical $1\sigma$ rms sensitivity of $\sim$0.5 mJy beam$^{-1}$ within the central 15$'$ $\times$ 15$'$ area.
We adopt a flux calibration uncertainty of $\sim$ 8\% at 1.15 mm, combining in quadrature the 5\% absolute uncertainty of the primary calibrator models with a representative rms calibration scatter of 6\% \citep{Perotto2020A&A}.

We cross-matched our NIKA2 map with published catalogs of background sources \citep{Stiele2011,Dey2023ApJ} and bright radio sources \citep{Walterbos1985} in M31. 
Sources identified as background galaxies or strong non-thermal radio emitters were masked prior to the SED fitting procedure to prevent bias in the derived dust properties.
This masking ensures that the subsequent analysis is dominated by dust thermal emission associated with M31.

A potential source of contamination in the 1.15 mm map is the CO(2–1) line emission, whose rest frequency (230.5 GHz) lies within the NIKA2 passband. 
The impact of this line contamination on the continuum measurements is assessed in Appendix~\ref{co_contamination}.

\subsection{Supplementary Data}\label{subsec:obs_sup}

\begin{table}[t]
\caption{Angular resolutions and adopted flux calibration
uncertainties of the data.}
\label{tab:data_summary}
\centering
\small
\setlength{\tabcolsep}{3pt}
\begin{tabular}{llccc}
\hline\hline
Telescope/Instrument & Band & FWHM & Cal. unc. & Ref. \\
                     &      & (arcsec) & (\%) & \\
\hline
\textit{Herschel}/PACS
        & 70~$\mu$m   & 5.6   & 5 & (1) \\
        & 100~$\mu$m  & 12.5  & 5 & (1) \\
        & 160~$\mu$m  & 13.3  & 5 & (1) \\
\textit{Herschel}/SPIRE
        & 250~$\mu$m  & 18.2  & 5.5  & (2) \\
        & 350~$\mu$m  & 24.5  & 5.5  & (2) \\
        & 500~$\mu$m  & 36.0  & 5.5  & (2) \\
JCMT/SCUBA-2
        & 450~$\mu$m  & 8.0   & 14 & (3) \\
        & 850~$\mu$m  & 14.0  & 6  & (3) \\
IRAM 30\,m/NIKA2
        & 1.15~mm    & 11.6  & 8  & (4) \\
\textit{Planck}/HFI
        & 217~GHz    & 299.4 & 0.54  & (5) \\
        & 353~GHz    & 289.2 & 1.24  & (5) \\
        & 545~GHz    & 280.8 & 10  & (5) \\
        & 857~GHz    & 259.8 & 10  & (5) \\
\hline
\end{tabular}
\tablefoot{
FWHM values refer to the input maps before resolution matching.
References specify the flux calibration uncertainties.
}
\tablebib{
(1) PACS Handbook and \citet{Balog2014ExA};
(2) SPIRE Handbook and \citet{Bendo2013MNRAS};
(3) \citet{Mairs2021AJ};
(4) \citet{Perotto2020A&A}, see Sect.~2.1;
(5) \citet{Planck2014VIII}.
}
\end{table}

M31 has been mapped by {\it Herschel} using the PACS instrument \cite{Poglitsch2010A&A}, the SPIRE instrument \cite{Griffin2010A&A}, and by {\it Planck} using the High Frequency Instrument \citep{Lamarre2010AA}.
In this work, we use the {\it Herschel} PACS 70/100/160 $\mu$m and SPIRE 250/350/500 $\mu$m data from \citet{Fritz2012A&A} and \citet{Groves2012MNRAS}.
The corresponding angular resolutions are 5.$''$6, 12.$''$5, 13.$''$3, 18.$''$2, 24.$''$5, and 36.$''$0, respectively.
We adopt an absolute flux calibration uncertainty of 5\% for the {\it Herschel}/PACS data, following the the PACS Handbook\footnote{PACS Handbook: \url{https://www.cosmos.esa.int/web/herschel/legacy-documentation-pacs}.}.
This uncertainty is primarily associated with the stellar calibrator models \citep{Balog2014ExA}. 
For the {\it Herschel}/SPIRE data, we adopt the conservative calibration uncertainty of 5.5\% recommended by the SPIRE Handbook\footnote{SPIRE Handbook:\url{https://www.cosmos.esa.int/web/herschel/legacy-documentation-spire}.}, obtained by linearly combining the 4\% absolute uncertainty in the Neptune calibration model and the 1.5\% relative calibration uncertainty derived from repeated measurements \citet{Bendo2013MNRAS}.
We retrieved the {\it Planck} High Frequency Instrument 217/353/545/857 GHz images\footnote{Based on observations obtained with {\it Planck} (http://www.esa.int/Planck), an ESA science mission with instruments and contributions directly funded by ESA Member States, NASA, and Canada.}, which are presented in \cite{Planck2015A&A}.
The {\it Planck} maps have an angular resolution of 4.$'$99, 4.$'$82, 4.$'$68, and 4.$'$33 at 217, 353, 545, and 857 GHz, respectively.
For the {\it Planck}/HFI data, we adopt calibration uncertainties of 0.54\%, 1.24\%, 10\%, and 10\% at 217, 353, 545, and 857~GHz, respectively, following \citet{Planck2014VIII}.

In addition to space-based observations, M31 has been mapped from the ground with JCMT-SCUBA2 at 450 and 850 $\mu$m by the JCMT Large Program HASHTAG\footnote{https://hashtag.astro.cf.ac.uk/DR1.html} \citep[HARP and SCUBA-2 High Resolution Terahertz Andromeda Galaxy Survey;][]{Li2020MNRAS,SmithApJS,Deng2025MNRAS} and the JCMT large PI project \citep{Jiao2025am31}.
We use the JCMT-SCUBA2 450 and 850 $\mu$m images from \cite{Jiao2025am31} in the subsequent SED analysis, which incorporated HASHTAG SCUBA2 data into the JCMT large PI project maps for the overlapping field of view.
These two maps have angular resolutions of $8''$ and $14''$,
respectively.
We adopt flux calibration uncertainties of 14\% at 450~$\mu$m and 6\% at 850~$\mu$m, based on the total-flux calibration uncertainties reported by \citet{Mairs2021AJ}.
The angular resolutions and adopted flux calibration uncertainties  for all data sets used are summarized in Table~\ref{tab:data_summary}.

\subsection{Combining NIKA2 Images with Planck Observations}\label{subsec:obs_com}

Ground-based bolometric (sub)millimeter continuum mapping observations of spatially extended sources are often subject to significant missing fluxes at large-scale \citep{Chapin2013MNRAS.430.2545C}. 
This limitation hampers accurate quantitative analyses. 
A common approach to recover this effect is to combine high-resolution ground-based maps with lower-resolution observations that preserve large-scale structures \citep[e.g.,][]{Liu2015ApJ...804...37L}.
In this work, we recover the missing large-scale flux in the NIKA2 1.15 mm image by combining it with archival {\it Planck} data \citep{Planck2015A&A} using the J-comb algorithm \citep{Jiao2022SCPMA}, following the procedure adopted by \cite{Lin2016ApJ,Lin2017ApJ} and \cite{Jiao2025am31}.

To generate the 260 GHz map that preserve large-scale structures, we made interpolations to the 260 GHz image based on the modified blackbody SED fitting to the {\it Planck} 217, 353, 545, and 857 GHz images. 
To obtain a sufficiently spatial frequency overlap for the NIKA2 observations, we applied an iterative deconvolution procedure to the interpolated {\it Planck}-based 260 GHz image following the method described in \citet{Jiao2025am31}.
The deconvolution was performed using the Lucy–Richardson algorithm \citep{Lucy1974}, which guarantees non-negative pixel values in the deconvolved image while conserving the total flux.
After the iterative deconvolution process, the interpolated {\it Planck} image was effectively deconvolved to a resolution of $60.0''$.
The NIKA2 260 GHz map and the deconvolved {\it Planck}-based 260 GHz image were then linearly combined in the Fourier domain using the J-comb algorithm \citep{Jiao2022SCPMA}.
This procedure preserves the high-resolution (11.$''$6) information from NIKA2 image while recovering the filtered extended emission.
The combined NIKA2 260 GHz image is shown in Figure \ref{fig:nika2_1mm}.

\begin{figure*}[ht!]
\vspace{-0.3cm}
\hspace{-0.1cm}
\begin{center}
\includegraphics[width=16.cm,]{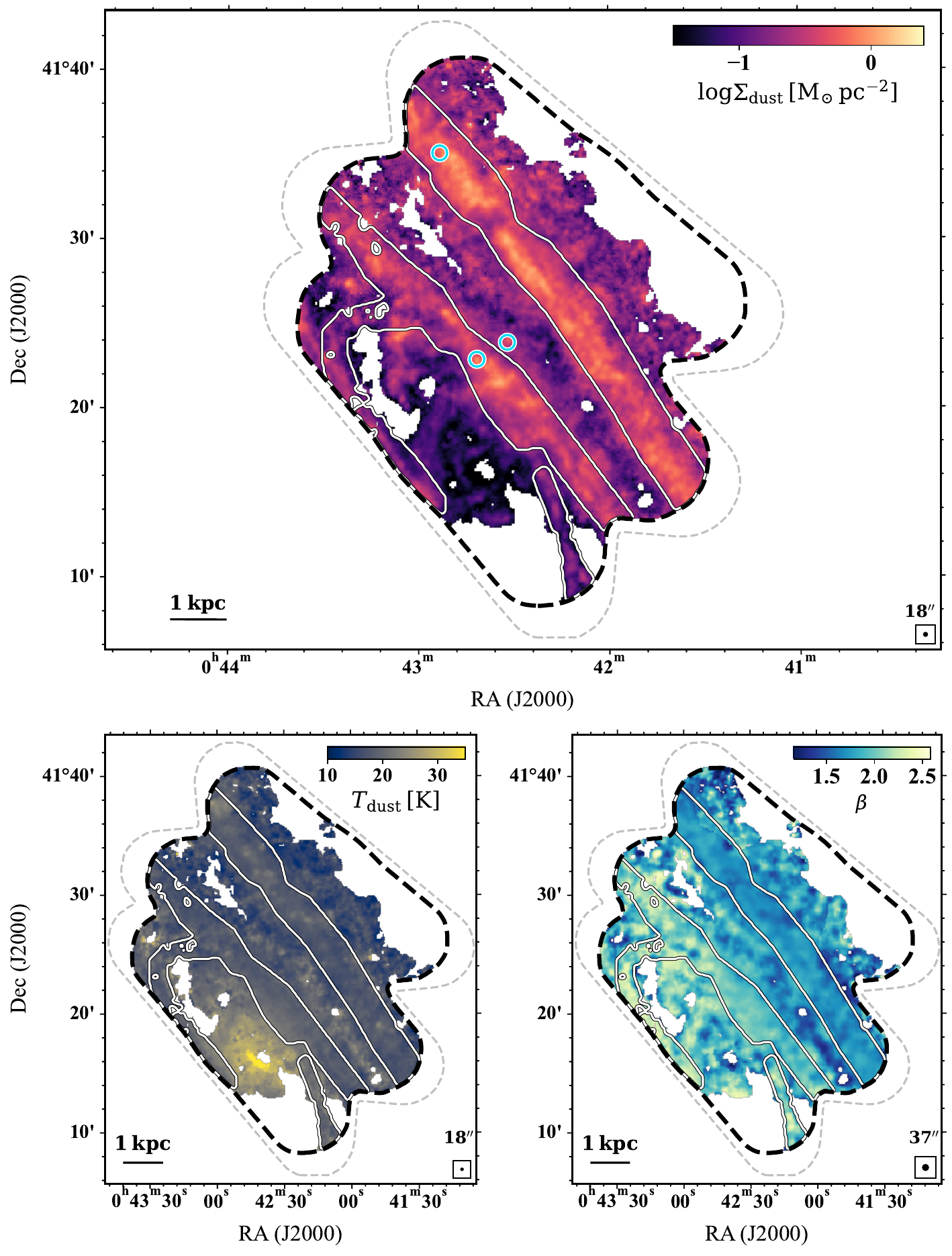}
\end{center}
\vspace{-0.3cm}
\caption{
The distribution of dust surface density (18$''$ angular resolution), temperature (18$''$ angular resolution), and $\beta$ (37$''$ angular resolution) obtained from the iterative SED fits using 250--1150~$\mu$m data.
The gray dashed line delineates the field of view of the new NIKA2 observations, while the black dashed contour corresponds to the 2 mJy beam$^{-1}$ sensitivity limit of the NIKA2 1.15 mm data.
The white contours trace the spiral-arm mask from \citet{Jiao2025am31}.
The cyan circles in the dust surface density map mark the three example positions used for the SED comparisons in Section \ref{subsec:T-beta}.}
\label{fig:sed_maps}
\end{figure*}

\subsection{SED fitting}\label{section:obs_sed}

We performed single-component, modified blackbody SED fitting to each pixel of the input images.
Before performing any SED fitting, we convolved all images to a common angular resolution of the largest telescope beam, and all images were re-gridded to have the same pixel size of 6$''$ per pixel.
In the least-squares fits, we weighted the data points by its corresponding noise level.
Under the modified blackbody assumption, the flux density $S_{\nu}$ at a certain observing frequency $\nu$ is expressed as
\begin{equation}\label{eq:sed_1}
S_{\nu} = \Omega_{m}B_{\nu}(T_{\mbox{\scriptsize dust}})(1-e^{-\tau_{\nu}}),
\end{equation}
where the dust surface density $\Sigma_{\mbox{\scriptsize dust}}$ can be approximated by
\begin{equation}\label{eq:sed_2}
\Sigma_{\mbox{\scriptsize dust}}=\tau_{\nu}/\kappa_{\nu},
\end{equation}
$B_{\nu}(T_{\mbox{\scriptsize dust}})$ is the Planck function at temperature $T_{\mbox{\scriptsize dust}}$, $\Omega_{m}$ is the solid angle, the dust opacity $\kappa_{\nu}=\kappa_{\mbox{\tiny{857 GHz}}}(\nu/857\,GHz)^{\beta}$, and $\beta$ is the dust emissivity index.
We adopted the dust opacity per unit mass at 857 GHz (350~$\mu$m) of 0.192 m$^{2}$ kg$^{-1}$ from \citet{Draine2003ARA&A}, consistent with previous M31 studies \citep[e.g.,][]{Smith2012ApJ,Jiao2025am31}. 
This assumed $\kappa$ normalization sets the absolute scale of the derived dust surface densities but is not independently constrained by the data.
The implications of this assumption are discussed at the beginning of Section~\ref{sec:dis}.
In the SED fitting, pixels with detections at a signal-to-noise ratio greater than 4 in fewer than three bands are masked.

Our procedure for iteratively deriving high-angular-resolution dust temperature and dust surface density maps follows \citet{Jiao2025am31}, with the addition of the NIKA2 260 GHz data.
We applied the same two-step SED fitting procedure to two wavelength configurations.
For the 70--1150~$\mu$m full-band fits, we first fitted the dust surface density, dust temperature, and $\beta$ simultaneously using the PACS 70, 100, and 160~$\mu$m data, the SPIRE 250, 350, and 500~$\mu$m data, the combined JCMT/SCUBA-2 450 and 850~$\mu$m maps, and the combined NIKA2 1.15~mm map.
This first step yields a $\beta$ map at an angular resolution of 37$''$.
We then fixed $\beta$ to the values obtained in the first step and fitted the PACS 70, 100, and 160~$\mu$m data, the SPIRE 250~$\mu$m data, the combined JCMT/SCUBA-2 450 and 850~$\mu$m maps, and the combined NIKA2 1.15~mm map to derive dust temperature and surface density maps at an angular resolution of 18$''$.

The analysis of \citet{Planck2015A&A} suggested that the 70, 100, and 160~$\mu$m emission may be more strongly associated with dust heated by recent star formation.
To reduce the possible influence of warmer dust components, we therefore repeated the SED fitting using only the 250, 350, 450, 500, 850, and 1150~$\mu$m data.
This long-wavelength configuration follows the same two-step procedure as the full-band fitting, except that the 70, 100, and 160~$\mu$m data are excluded from the fit and retained only as upper limits.
The two wavelength configurations yield broadly similar spatial distributions and radial trends in the fitted dust properties.
We therefore adopt the 250--1150~$\mu$m results as the focused results presented in Section~\ref{sec:res} and used for the main discussion in Section~\ref{sec:dis}, while the corresponding full-band SED fitting results are provided in Appendix~B for comparison.

\section{Results}\label{sec:res}

\subsection{Surface Density, Temperature, and Dust Opacity Index Distribution Maps from SED Fitting}\label{subsec:res_t_n}

Figure~\ref{fig:sed_maps} presents the results of our iterative SED fitting procedure using 250--1150~$\mu$m data: the dust surface density ($\Sigma_{\mbox{\scriptsize dust}}$) map at 18$''$ resolution, the dust temperature ($T_{\mbox{\scriptsize dust}}$) map at 18$''$ resolution, and the dust emissivity index ($\beta$) map at 37$''$ resolution, derived from pixel-by-pixel modified blackbody fitting.
The gray dashed line delineates the field of view of the new NIKA2 observations, while the black dashed contour corresponds to the 2 mJy beam$^{-1}$ sensitivity limit of the NIKA2 1.15 mm data.
In the final maps shown in Figure~\ref{fig:sed_maps}, pixels with fewer than three bands detected at signal-to-noise ratios greater than 4$\sigma$ are masked.
This threshold was adopted as an empirical quality cut for removing the low-signal regions, as some regions remained visibly noisy even at a $3\sigma$ threshold. 
Overall, after incorporating the NIKA2 1.15 mm data into the SED fitting, the derived dust property distributions are consistent with those reported in \cite{Jiao2025am31}.

\begin{figure}
\vspace{0.1cm}
\includegraphics[width=9.cm]{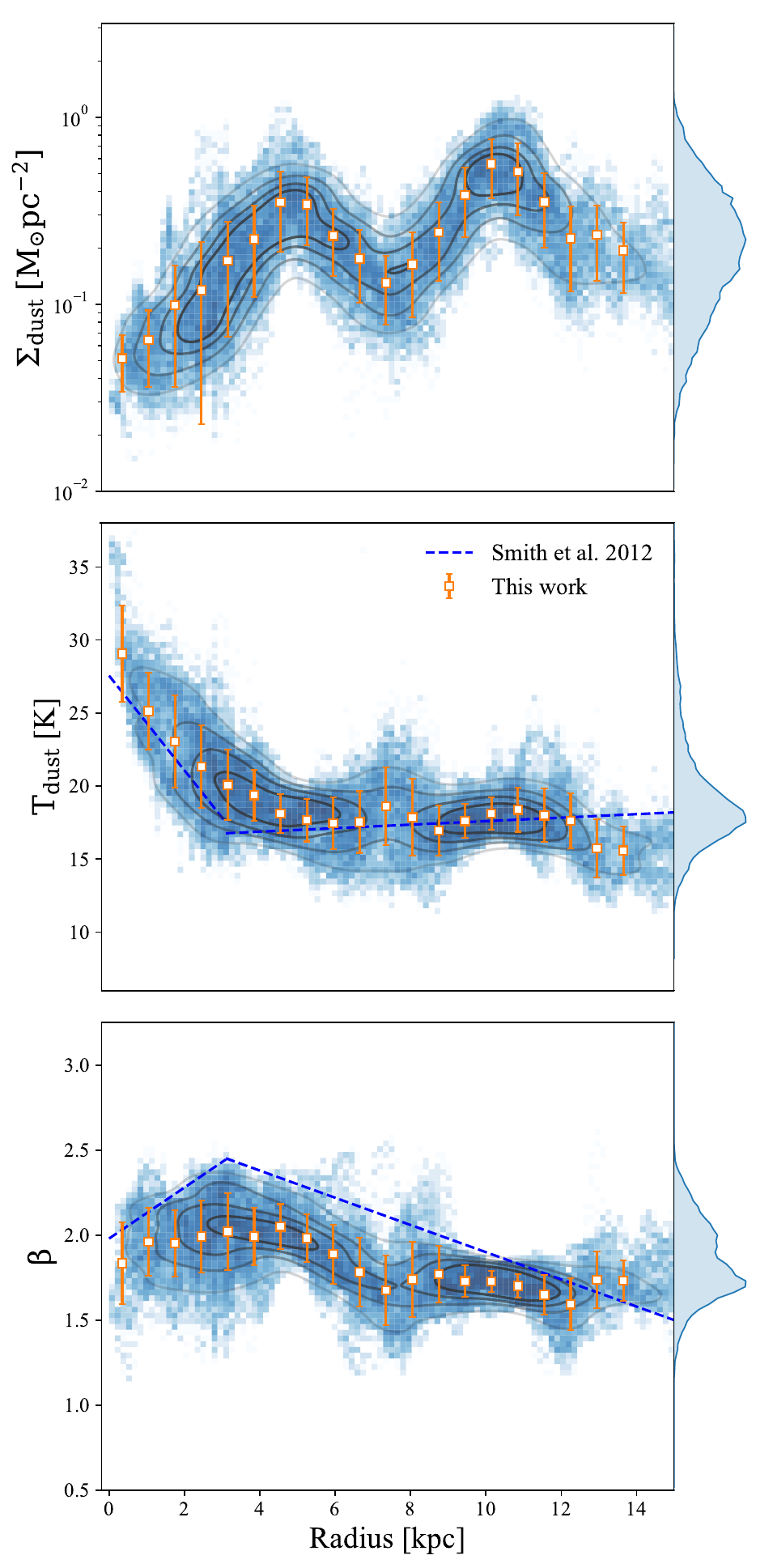}
\caption{
Results from the 250--1150~$\mu$m band iterative SED fits for each pixel plotted vs. radius.
The blue points show the fitted value for each pixel, while the yellow squares represent averages in 700 pc bins.
The corresponding yellow error bars represent the standard deviation within each bin.
The dashed blue lines represent the fitted relationships from the HELGA II \citep{Smith2012ApJ}.
The histograms on the right side show the dust surface density, dust temperature, and $\beta$ distribution of the map region shown in Figure \ref{fig:sed_maps}.}
\label{fig:ori_sed_high_nh2}
\end{figure}

The dust surface density, $\Sigma_{\mbox{\scriptsize dust}}$, map (top panel of Figure~\ref{fig:sed_maps}) reveals a highly structured and filamentary morphology, with $\Sigma_{\rm dust}$ ranging from $\sim$0.05 to 1 M$_{\odot}$ pc$^{-2}$ across the target region.
The highest $\Sigma_{\mbox{\scriptsize dust}}$ values trace the prominent large-scale structures, mainly the 5 kpc ring and 10 kpc ring, while the dust surface density in the galactic center is more than an order of magnitude lower.
In contrast, the inter-ring regions exhibit significantly lower surface densities, although a number of GMC-scale structures are still apparent.
In addition, this map also reveals several relatively diffuse structures extending from the spiral arms into the inter-ring regions, similar to features identified in other nearby galaxies using other molecular gas tracers \citep[e.g.,][]{Koda2009ApJ,Colombo2014ApJ,Duarte-Cabral2017,Meidt2023ApJ,Bazzi2026}.

The dust temperature map (bottom left panel of Figure~\ref{fig:sed_maps}) displays comparatively smoother spatial variations.
Across most of the region, $T_{\rm dust}$ ranges from $\sim$15 to $\sim$20 K, while the central $\sim$3 kpc of M31 shows systematically elevated temperatures reaching up to $\sim$30 K.
The enhanced dust temperatures in the central $\sim$3 kpc are likely driven by the strong interstellar radiation field associated with the bulge-dominated stellar population \citep{Viaene2017}.
Previous studies have demonstrated that dust heating in the inner regions of M31 correlates more strongly with stellar mass surface density than with star formation tracers, indicating that evolved stars dominate the heating budget \citep[e.g.,][]{Smith2012ApJ,Draine2014ApJ,Viaene2017}.

The dust emissivity index $\beta$ map (bottom right panel of Figure~\ref{fig:sed_maps}) exhibits relatively modest spatial variations, with $\beta$ values ranging from $\sim$1.5 to $\sim$2.1 and an average value of $\sim$ 1.8. 
This is broadly consistent with measurements of the diffuse ISM in the Milky Way \citep[e.g.,][]{Planck2011A&A,Draine2011}.
On large scales, $\beta$ shows a mild decreasing trend from the galactic center toward the 10 kpc ring regions and does not display a strong spatial correspondence with the dense spiral arms or rings.
Although the overall $\beta$ variation is limited, the $\beta$ distribution is not entirely uniform and forms coherent structures on several-hundred-pc scale.

\begin{figure*}[ht!]
\vspace{-0.3cm}
\hspace{-0.1cm}
\includegraphics[width=18.2cm,]{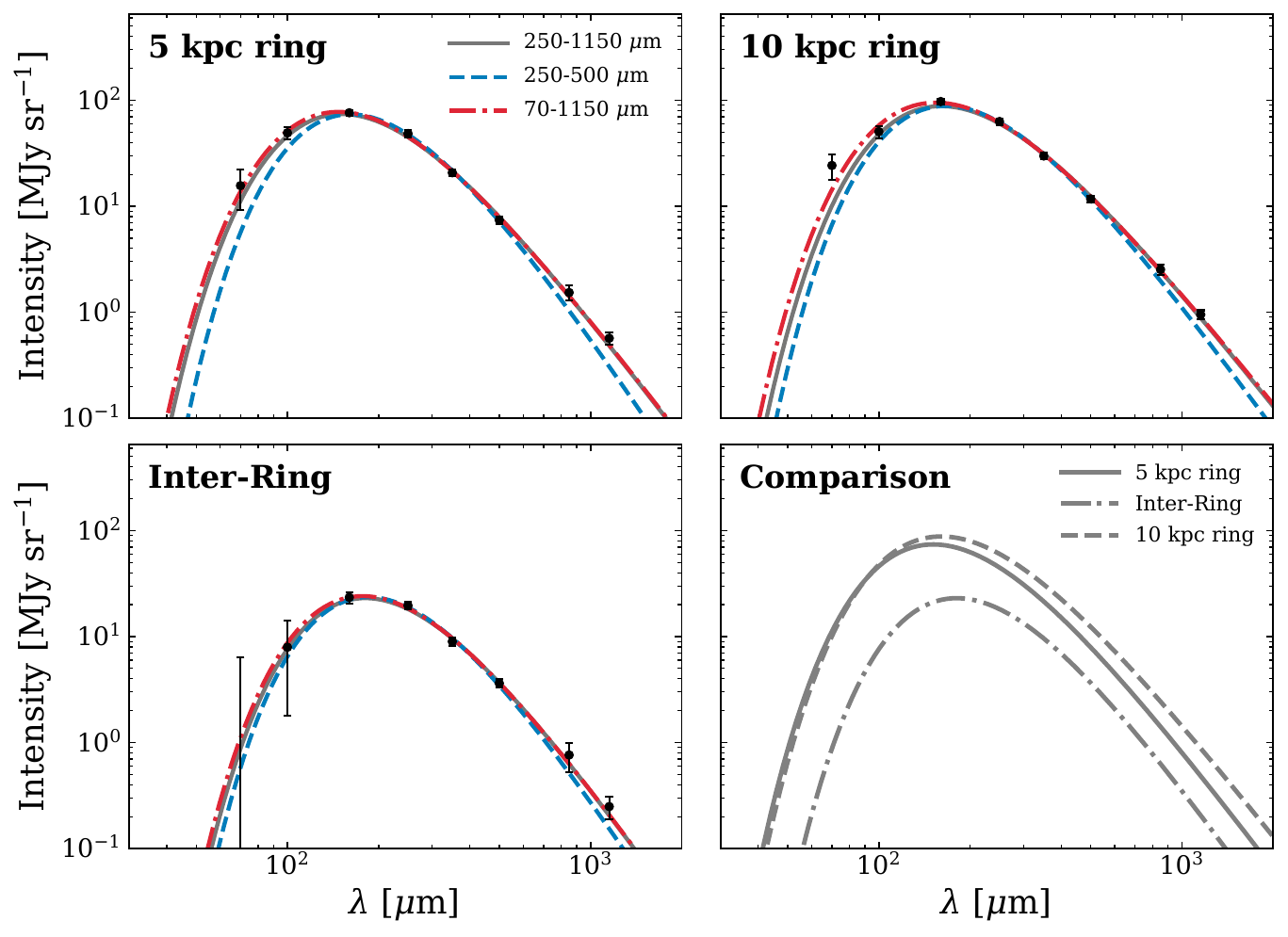}
\caption{ 
Examples of dust SED fitting for the 5 kpc ring (upper left), inter-ring region (lower left), and 10 kpc ring (upper right). 
The black points with error bars indicate the observed intensities and uncertainties.
The gray solid curves show fits to the 250--1150~$\mu$m data, the blue dashed curves show fits to the 250--500~$\mu$m data, and the red dash-dotted curves show fits to the 70--1150~$\mu$m data.
The lower-right panel compares the 250--1150~$\mu$m models for the 5 kpc ring (solid), 10 kpc ring (dashed), and inter-ring region (dash-dotted).
The representative positions (RA, Dec; J2000) are (00:42:41.58; +41:22:53.39)}, (00:42:31.97; +41:23:53.19), and (00:42:53.24; +41:35:05.57) for the 5 kpc ring, inter-ring region, and 10 kpc ring, respectively.
\label{fig:sed_example}
\end{figure*}

\subsection{Dust Property Trends with Galactocentric Radius} \label{subsec:trends}

Figure~\ref{fig:ori_sed_high_nh2} presents the radial distributions of $\Sigma_{\rm dust}$, $T_{\rm dust}$, and $\beta$ as functions of galactocentric radius.
Each blue point represents the fitted value for an individual pixel, while the yellow squares denote the average values in 700 pc radial bins.
The dashed blue curves correspond to the radial relations derived in the HELGA II study by \citet{Smith2012ApJ}.
The histograms on the right-hand side show the overall distributions of $\Sigma_{\rm dust}$, $T_{\rm dust}$, and $\beta$ within the mapped region.

For the selected target region, the dust surface density exhibits a more pronounced radial structure than that reported for the entire disk of M31 \citep[e.g.,][]{Smith2012ApJ,Whitworth2019MNRAS,Jiao2025am31}.
$\Sigma_{\rm dust}$ increases from the galactic center to a first enhancement near the $\sim$5 kpc ring and reaches a more prominent peak at the well-known 10 kpc ring.
Between these two ring structures, the surface density decreases significantly.
Beyond the 10 kpc ring, $\Sigma_{\rm dust}$ gradually declines toward larger radii.

The dust temperature shows a systematic radial gradient that is broadly consistent with previous {\it Herschel}-based studies \citep{Smith2012ApJ}.
$T_{\rm dust}$ reaches its highest values within the central $\sim$3 kpc (approaching $\sim$30 K) and declines rapidly to $\sim$17–18 K at radii of $\sim$5 kpc.
At larger radii, the temperature profile becomes relatively flat, with only mild variations toward the outer disk.

The emissivity index $\beta$ exhibits a comparatively shallow but coherent radial trend.
Within the inner $\sim$7 kpc, the median $\beta$ value is 2.03, while beyond $\sim$7 kpc it decreases to a median value of 1.71.
Compared to the results of \citet{Smith2012ApJ}, our $\beta$ profile appears flatter across the disk.
After incorporating the longer-wavelength data into the SED fitting, we derive an average $\beta$ value of 1.8 for the target region, which is lower than values reported in previous shorter-wavelength-only studies \citep{Smith2012ApJ,Draine2014ApJ,Whitworth2019MNRAS}.
The impact of including the longer-wavelength data in the SED fitting is discussed further in Section~\ref{subsec:T-beta}.

\section{Discussion}\label{sec:dis}

\begin{figure}
\centering
\includegraphics[width=7.7cm]
{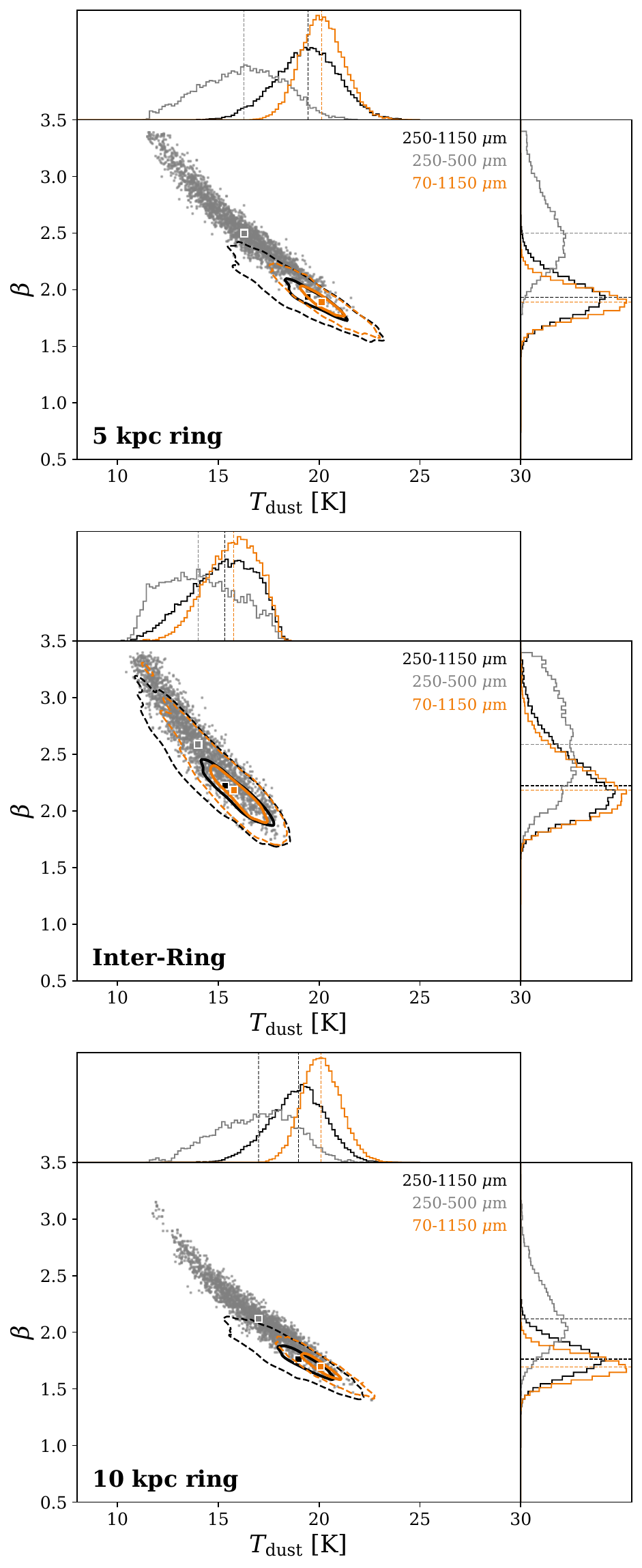}
\vspace{-0.4cm}
\caption{
Joint posterior distributions of $T_{\rm dust}$ and $\beta$ derived from SED fitting with different band coverages.
The black and orange contours represent fits using the 250-1150 $\mu$m band and 70-1150 $\mu$m band respectively. 
Solid and dashed contours correspond to the 0.5 and 0.9 density levels. 
The case using 250-500 $\mu$m bands is shown as gray scatter points. 
Marginal posterior distributions of $T_{\rm dust}$ and $\beta$ are shown in the top and right panels. 
Colored dashed lines and square markers indicate the median values of each distribution.}
\label{fig:T-beta}
\end{figure}

The dust surface-density map and radial profile in Figures~\ref{fig:sed_maps} and \ref{fig:ori_sed_high_nh2} depend on the adopted opacity,
$\kappa_{\mbox{\tiny{857 GHz}}} =0.192\,{\rm m^2\,kg^{-1}}$ \citep{Draine2003ARA&A}.
In our single-component modified-blackbody model, $\Sigma_{\mbox{\scriptsize dust}}=\tau_{\nu}/\kappa_{\nu}$,
with $\kappa_{\nu}=\kappa_{\mbox{\tiny{857 GHz}}} (\nu/857\,GHz)^{\beta}$.
The SED fit constrains the optical depth and an effective $\beta$, but cannot independently determine the absolute opacity normalization and dust surface density.
In particular, the spatial variation in fitted $\beta$ does not establish that the adopted $\kappa_{\mbox{\tiny{857 GHz}}}$ is
appropriate at every position.
At fixed fitted optical depth, a uniform change in $\kappa_{\mbox{\tiny{857 GHz}}}$ would rescale the derived surface densities and dust masses.
For example, the far-infrared power-law approximation to the THEMIS dust model \citep{Jones2017}, as presented by \citet{Galliano2018}, gives $\kappa_{\mbox{\tiny{857 GHz}}}\simeq0.35\,{\rm m^2\,kg^{-1}}$.
Replacing only our adopted opacity with this value, while retaining the fitted optical depths and $\beta$ values, would reduce the derived surface densities and masses by approximately a factor of two.
If the true opacity instead varies spatially, the contrast of structures in the surface-density map and the shape of its radial profile could change as well.
We therefore regard the reported dust surface densities as estimates conditional on the adopted opacity prescription.

\subsection{Effects of Wavelength Coverage on the $T_{\mbox{\scriptsize dust}}$ -- $\beta$ Degeneracy}\label{subsec:T-beta}

In single-temperature modified blackbody SED fitting (Section~\ref{section:obs_sed}), shorter-wavelength observations ($<$0.5 mm) primarily constrain the dust temperature $T_{\rm dust}$, while longer-wavelength data provide stronger leverage on the emissivity index $\beta$.
In the Rayleigh–Jeans limit, $B_{\nu}(T_{\rm dust}) \propto \nu^2$, such that the spectral slope becomes less sensitive to temperature and more directly probes $\beta$.
In M31, except for the central $\sim$3 kpc region, the typical dust temperature is $\sim$18 K.
At such temperatures, none of the {\it Herschel} bands are in the Rayleigh–Jeans approximation.
Consequently, reliable constraints on $\beta$ require an accurate determination of $T_{\rm dust}$.
In the absence of observations in the Rayleigh-Jeans limit, $T_{\rm dust}$ and $\beta$ are known to exhibit strong degeneracy in modified blackbody SED fitting \citep{Desert2008,Shetty2009ApJa,Shetty2009ApJb}. 
This $T_{\rm dust}$--$\beta$ degeneracy has also been reported in M31 based on {\it Herschel} observations \citep{Smith2012ApJ}. 
Restricting the SED fitting to wavelengths $\lambda\geq250~\mu$m reduces the sensitivity to warmer dust components that contribute more strongly at shorter wavelengths.
This may mitigate the risk of an artificial inverse relation between $T_{\rm dust}$ and $\beta$ caused by temperature mixing in single-temperature fits \citep{Shetty2009ApJb}.
The resulting relation may therefore better reflect the properties of the colder dust.

To improve the long-wavelength coverage of the SED, we incorporate SCUBA-2 850~$\mu$m and NIKA2 1.15 mm data.
To quantify the impact of wavelength coverage, we conduct a focused test at three example positions in the 5~kpc ring, the inter-ring region, and the 10~kpc ring, following a procedure similar to that described by \citet{Feng2026ApJ}. 
Their respective coordinates are $(\mathrm{RA},\mathrm{Dec}; J2000)$ = (00:42:41.58; +41:22:53.39), (00:42:31.97; +41:23:53.19), and (00:42:53.24; +41:35:05.57), while these positions are marked by cyan circles in the dust surface density map in Figure~\ref{fig:sed_maps}.

We carry out SED fitting at these positions under three different configurations: (a) 250, 350, 450, 500, 850~$\mu$m, and 1150~$\mu$m; (b) 70-1150~$\mu$m; and (c) 250, 350, and 500~$\mu$m bands.
The comparison between configuration (a) and (b) allows us to assess the effect of including the shorter-wavelength data, while the comparison between configuration (a) and (c) tests the effect of including the SCUBA-2 850~$\mu$m and NIKA2 1.15~mm measurements in the SED fitting.
In each case, we employ Markov Chain Monte Carlo (MCMC) sampling to derive posterior distributions for $T_{\rm dust}$ and $\beta$. 
Figure~\ref{fig:sed_example} illustrates the SED fits within a single panel for each position, with the observed intensities and their uncertainties shown as black points with error bars.
The final panel compares the SEDs of the three regions for configuration (a).
The fitted $\Sigma_{\rm dust}$, $T_{\rm dust}$, and $\beta$, together with their uncertainties, are listed in Table~\ref{tb_sed_results}.

\begin{table}
\caption{SED fitting results at the three selected positions using different wavelength ranges.}
\label{tb_sed_results}
\centering
\small
\begin{tabular}{lcccc}
\hline\hline
Position & Wavelength & $\Sigma_{\rm dust}$ & $T_{\rm dust}$ & $\beta$ \\
& ($\mu$m) & ($M_\odot\,{\rm pc}^{-2}$) & (K) & \\
\hline
5 kpc ring
& 250--1150 & $0.420_{-0.081}^{+0.117}$
& $19.4_{-1.6}^{+1.5}$ & $1.93_{-0.16}^{+0.18}$ \\
& 70--1150 & $0.386_{-0.056}^{+0.067}$
& $20.1_{-1.0}^{+1.1}$ & $1.89_{-0.13}^{+0.13}$ \\
& 250--500 & $0.670_{-0.183}^{+0.379}$
& $16.3_{-2.2}^{+2.0}$ & $2.50_{-0.28}^{+0.38}$ \\
\hline
Inter-ring
& 250--1150 & $0.352_{-0.089}^{+0.183}$
& $15.3_{-1.8}^{+1.6}$ & $2.22_{-0.24}^{+0.34}$ \\
& 70--1150 & $0.326_{-0.068}^{+0.119}$
& $15.8_{-1.4}^{+1.2}$ & $2.18_{-0.21}^{+0.28}$ \\
& 250--500 & $0.465_{-0.165}^{+0.301}$
& $14.0_{-1.9}^{+2.2}$ & $2.59_{-0.41}^{+0.45}$ \\
\hline
10 kpc ring
& 250--1150 & $0.646_{-0.119}^{+0.165}$
& $19.0_{-1.5}^{+1.3}$ & $1.76_{-0.14}^{+0.14}$ \\
& 70--1150 & $0.563_{-0.077}^{+0.088}$
& $20.1_{-0.9}^{+1.0}$ & $1.69_{-0.10}^{+0.11}$ \\
& 250--500 & $0.850_{-0.203}^{+0.419}$
& $17.0_{-2.2}^{+1.8}$ & $2.12_{-0.23}^{+0.32}$ \\
\hline
\end{tabular}
\end{table}

We first compare configurations (a) and (b) to examine the effect of including the shorter-wavelength measurements.
In all three examples, the 70--1150~$\mu$m fits yield higher dust temperatures and lower $\beta$ values than the 250--1150~$\mu$m fits.
The inclusion of the shorter-wavelength data has a more noticeable effect on the inferred temperatures, while the $\beta$ values remain relatively stable and their posterior distributions overlap substantially (Figure~\ref{fig:T-beta}).
This behavior may reflect a larger contribution from warmer dust to the shorter-wavelength emission, which shifts the single-temperature fit toward higher temperatures.

Comparing configurations (a) and (c), including the 850 and 1150~$\mu$m measurements flattens the fitted millimeter slope and leads to lower inferred $\beta$ values.
Figure~\ref{fig:T-beta} shows that these two additional measurements also narrow the allowed parameter ranges.
When the 850 and 1150 $\mu$m data are used, both the fitted $T_{\rm dust}$ and $\beta$ have smaller uncertainties and the median $\beta$ value decreases relative to the {\it Herschel}-only configuration.
However, Fig.~\ref{fig:T-beta} also shows that the posterior contours remain elongated, especially for the cold inter-ring region, with higher temperatures corresponding to lower $\beta$ values.
This shape is similar to the fitting-induced anticorrelation discussed by \citet{Shetty2009ApJa}.
Thus, the inclusion of 850~$\mu$m and 1150~$\mu$m data may reduce but can not remove the $T_{\rm dust}$--$\beta$ degeneracy. 

\begin{figure}
\vspace{0.01cm}
\includegraphics[width=9.cm]{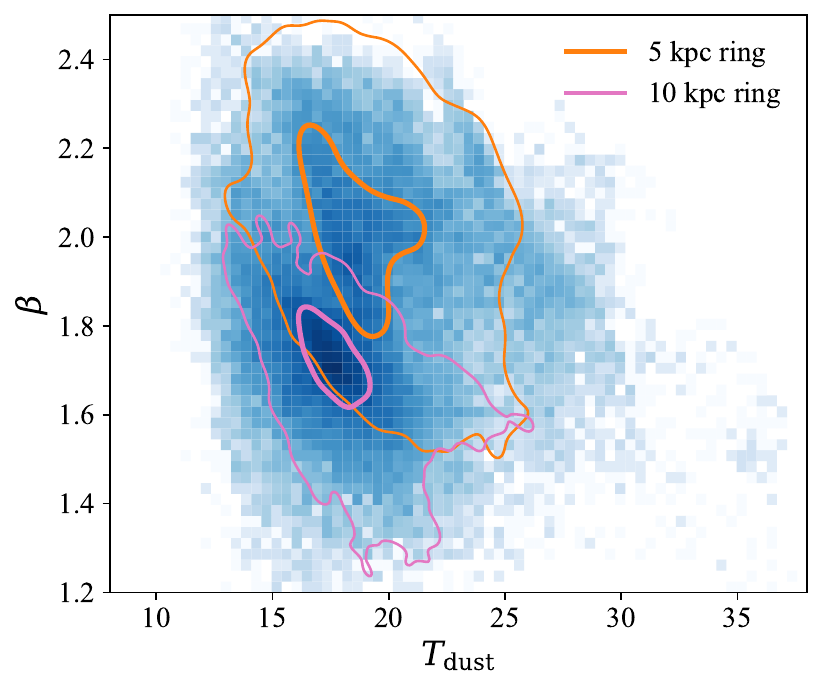}
\caption{
Two-dimensional histograms of $T_{\rm dust}$ and $\beta$ for mapping region. 
Contours for the 5 kpc (orange) and 10 kpc (pink) rings highlight the corresponding distributions, with thin and thick lines representing the 1\% and 50\% levels, respectively.
}
\label{fig:T_beta_distribution}
\end{figure}

To further assess this issue across the full mapped region, we examine the pixel-by-pixel distributions of $T_{\rm dust}$ and $\beta$ (Figure \ref{fig:T_beta_distribution}).
The full sample does not show a strong global anti-correlation, suggesting that the longer-wavelength data help to reduce the apparent coupling between the two parameters.
Nevertheless, within the 5 kpc and 10 kpc ring subsamples, a weak negative correlation remains.
This residual correlation is expected because much of the target region has low dust temperatures of $\sim$18 K, for which the current wavelength coverage still does not well sample the Rayleigh–Jeans limit.
The two ring samples also occupy distinct regions in the $T_{\rm dust}$--$\beta$ plane, particularly in $\beta$, which is further discussed in Section \ref{subsec:beta_r}.
Further improvement may be achieved by incorporating higher-resolution measurements at longer wavelengths, for example, GMC-scale observations at 3 mm with GBT/MUSTANG-2 \citep{Dicker2014JLTP}, which would extend the leverage deeper into the Rayleigh–Jeans limit.

As discussed above, including the SCUBA-2 850~$\mu$m and NIKA2 1.15 mm data systematically yields lower $\beta$ values than fits based only on {\it Herschel} observations. 
The elevated millimeter flux densities relative to {\it Herschel}-only extrapolations naturally flatten the long-wavelength SED slope and hence lead to smaller fitted $\beta$.
This suggests that the $\beta$ inferred from SED fitting in this region might be lower than previously inferred from shorter wavelength only analyses \citep{Smith2012ApJ,Draine2014ApJ,Whitworth2019MNRAS}.
Nevertheless, alternative explanations for the enhanced long-wavelength emission should be considered.
One possibility is contamination from free--free emission.
Global SED modeling of M31 indicates that the contribution of free--free emission at wavelengths $\sim$1 mm is expected to be modest \citep{Battistelli2019ApJ}.
Although it is unlikely to dominate the observed excess, higher-resolution radio continuum observations, such as those from the Local Group L-Band Survey \citep[LGLBS;][]{Pingel2024ApJ,Koch2025ApJS}, will help to better quantify and subtract any residual free--free contribution at millimeter wavelengths.

\subsection{Comparing Dust Emissivity Index in 5 kpc Ring and 10 kpc Ring}\label{subsec:beta_r}

Within the mapped region, two prominent large-scale structures with enhanced dust surface density are evident, e.g., the 5 kpc ring and the 10 kpc ring, both of which have been identified in other tracers \citep[e.g.,][]{Nieten2006A&A,Gordon2006ApJ}.
These ring structures are clearly visible in the radial profile of $\Sigma_{\rm dust}$ ( top panel of Figure \ref{fig:ori_sed_high_nh2}), where the surface density exceeds $\sim$1 M$_{\odot}$ pc$^{-2}$.
While the two rings exhibit similar dust temperatures of $\sim$18 K, the emissivity index $\beta$ tends to be higher in the inner (5 kpc) ring than in the outer (10 kpc) ring.

\begin{figure}
\vspace{0.01cm}
\includegraphics[width=8.7cm]{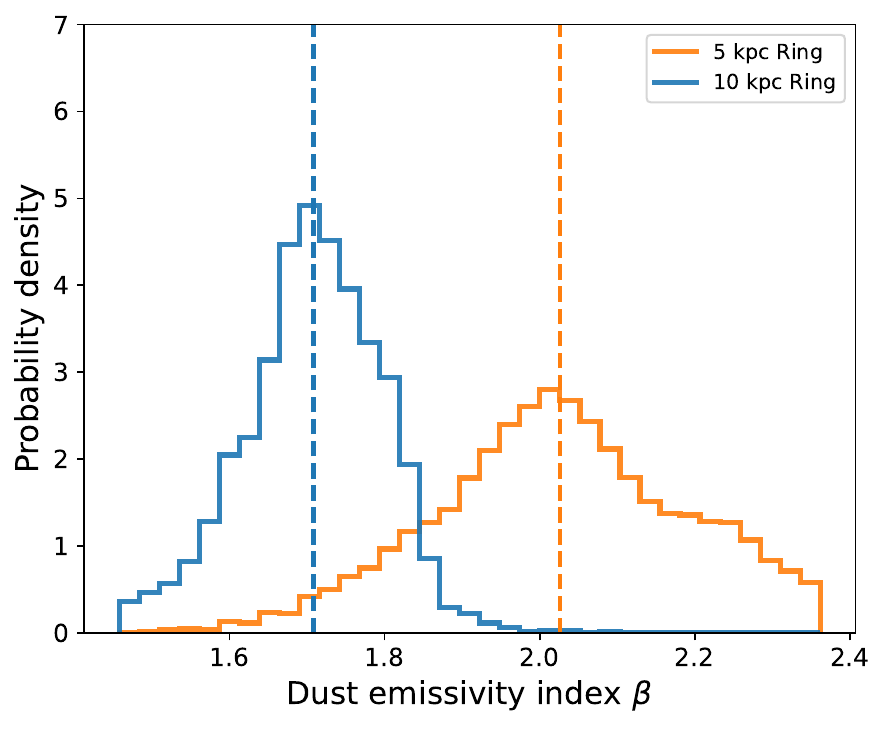}
\caption{
Distribution of the dust emissivity index $\beta$, separated into 5 kpc ring and 10 kpc ring. 
The histograms are normalized to probability density to account for the different number of pixels in each bin. 
Dashed vertical lines indicate the median values.
}
\label{fig:beta_distribution}
\end{figure}

To assess whether this difference is statistically significant, we perform a systematic comparison of $\beta$ values within the two rings, focusing on regions with the highest signal-to-noise ratios.
We adopt the spiral-arm mask from \cite{Jiao2025am31} to exclude low signal-to-noise areas and restrict the analysis, which is based on dust tracers and is shown in Figure \ref{fig:sed_maps}.
The masked $\beta$ map is then divided into two radial samples based on galactocentric distance.
Figure \ref{fig:beta_distribution} presents the normalized $\beta$ distributions for the 5 kpc ring (orange histogram) and the 10 kpc ring (blue histogram). 
The distributions are normalized to allow direct comparison independent of the different number of pixels.
Dashed vertical lines mark the median values (2.03 for the 5 kpc ring and 1.71 for the 10 kpc ring, respectively).
The 5 kpc ring exhibits a broader and more dispersed $\beta$ distribution with a higher median value, whereas the 10 kpc ring shows a comparatively narrower distribution.
Because the $6''$ pixels oversample the $37''$ resolution of the $\beta$ map, we performed the Mann--Whitney U test \citep{Mann47} on median $\beta$ values measured in non-overlapping $72''$ spatial blocks. 
We repeated the test for all possible pixel offsets of the block grid to reduce sensitivity to its placement. 
For all grid placements, the two ring samples remain significantly different, with a conservative maximum $p$-value of $4.7\times10^{-15}$ and a minimum rank-biserial correlation coefficient of $r_{\rm rb}=0.989$, where positive values indicate higher $\beta$ in the 5 kpc ring.

Previous studies of M31 have reported systematic radial variations in $\beta$, with generally higher values in the inner regions and lower values toward the outer disk \citep[e.g.,][]{Smith2012ApJ,Draine2014ApJ,Whitworth2019MNRAS,Athikkat-Eknath2022MNRAS}.
The distinct $\beta$ distributions between the 5 kpc and 10 kpc rings found in this work are consistent with these trends.
\cite{Draine2014ApJ} found that their silicate–graphite–PAH dust model systematically overpredicts the 500 $\mu$m emission at radii of $\sim$2–6 kpc, suggesting possible spatial variations in dust optical properties in the inner disk.
The broader fitted $\beta$ distribution in the 5 kpc ring may reflect greater variations in dust properties or local environmental conditions.
However, the lower surface brightness of the 5 kpc ring may also lead to larger fitting uncertainties, contributing to the broader observed distribution.

Besides variations in dust optical properties, the dust spectral index correlates with the mean starlight heating intensity, $\langle U \rangle$, which can affect the dust compositions.
Here, $\langle U \rangle$ is defined as a dimensionless scaling factor of the interstellar radiation field relative to that in the solar neighbourhood \citep{Draine2007SINGS}.
Higher $\langle U \rangle$ can promote photo-processing and mantle erosion, potentially enriching the dust mixture in silicate components and steepening the emissivity slope \citep[e.g.,][]{Jones2013,Ysard2016}.
Observational studies support such environmentally driven variations \citep[e.g.,][]{Compiegne2011,Galliano2018,Pantoni2026}.
In M31, $\langle U \rangle$ peaks in the central region, reaching values of $\sim$35, and declines to $\sim$0.25 at $R\sim20$ kpc \citep{Draine2014ApJ}.
The systematically higher fitted $\beta$ values in the inner 5 kpc ring may therefore be linked to enhanced $\langle U \rangle$ in this region.

Differences in $\beta$ could also arise from changes in the dust size distribution \citep[e.g.,][]{Ormel2011,Schnee2014MNRAS}.
In the Rayleigh regime $(a \ll \lambda/2\pi)$, the absorption opacity is largely independent of grain size \citep{Bohren1983}.
However, once grains grow to sizes comparable to the observed wavelength ($a$ $\gtrsim \lambda/2\pi$), both $\kappa_{\mathrm{abs}}$ and $\beta$ become sensitive to grain size as the scattering behavior transitions to the Mie or geometric regime \citep{Draine2006ApJ}.
Interpreting the observed enhancement in $\beta$ purely in terms of grain growth would require characteristic grain sizes of at least several tens of $\mu$m, significantly larger than the typical sub-micron ($\sim0.1~\mu$m) grains in the diffuse ISM \citep[e.g.,][]{Draine2011}.
In practice, however, the relationship between grain growth and $\beta$ is more complex.
Grain growth is expected to be accompanied not only by changes in grain size but also by modifications in grain structure (e.g., the formation of ice mantles), which can also affect the observed $\beta$ \citep[e.g.,][]{Jones2013,Ysard2016}.
In addition, the presence of very cold dust components and temperature mixing along the line of sight may influence the inferred SED and weaken direct constraints on large grains \citep[e.g.,][]{Li1999ApJ}.
Given these complications, grain size variations may contribute to the observed $\beta$ differences but are unlikely to provide a unique explanation for the observed $\beta$ in M31.

\begin{figure}
\vspace{0.01cm}
\includegraphics[width=8.7cm]{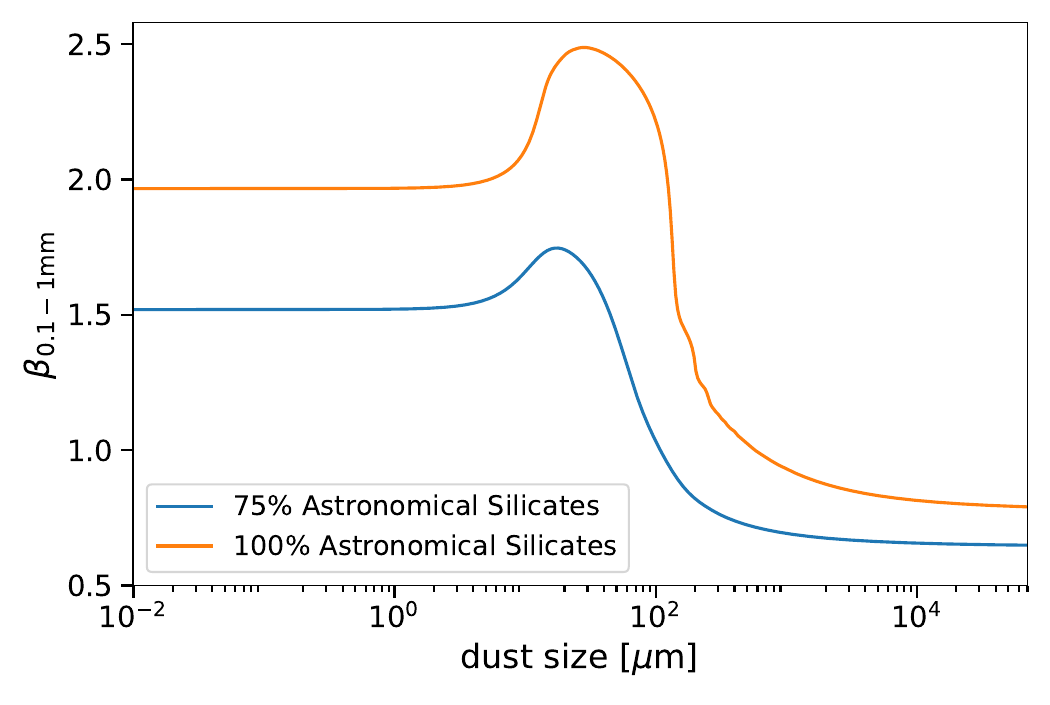}
\caption{
$\beta_{0.1-1{\rm mm}}$ profile with different averaged grain sizes.
}
\label{fig:T_lam}
\end{figure}

To further illustrate how variations in grain size and composition may affect the observed $\beta$, we computed absorption opacities using Mie theory with the \texttt{dsharp\_opac} package \citep{Birnstiel2018ApJ}.
We adopted the dust model based on the dust constants of Graphite from \citet{Laor1993ApJ} and Astronomical Silicates from \citet{Draine2003ARA&A}.
Within this framework, we explored two dust mixtures: (i) 75.1\% of the dust mass is contributed by Astronomical Silicates and 24.9\% by Graphite, following \citet{Draine2014ApJ}; (ii) 100\% of Astronomical Silicates.
The grain size distribution is assumed to follow a power law,
\begin{equation}\label{eq:pl}
N(a) \propto a^{-p},
\end{equation}
where $N(a)$ is the number density of dust grains with size $a$. 
We adopted the classical distribution of $p = 3.5$ as measured by \citet{Mathis1977ApJ} in the ISM from infrared extinction.
Figure~\ref{fig:T_lam} presents the dependence of the spectral index $\beta_{0.1-1{\rm mm}}$, defined from the logarithmic slope of the absorption opacity over 0.1–1 mm, as a function of averaged grain size.
As the astronomical silicates mass fraction increases, the resulting $\beta$ becomes larger.
The resulting $\beta$–size relation is non-monotonic, as reported in previous studies \citep[e.g.,][]{Birnstiel2018ApJ}, and exhibits a local maximum at grain sizes of order $\sim$10~$\mu$m.
These results indicate that variations in $\beta$ may arise from a combination of changes in dust optical properties, composition, and grain size distribution.


\section{Conclusions}\label{sec:con}

We present new IRAM 30m NIKA2 1.15 mm observations of M31 covering the 5 kpc and 10 kpc rings along the minor axis. 
By combining these data with {\it Herschel} and JCMT SCUBA-2 observations, we perform pixel-by-pixel modified blackbody fitting to derive maps of dust surface density, temperature, and emissivity index.
Our main findings are:

\begin{enumerate}
\item The inclusion of 850 $\mu$m and 1.15 mm data helps reduce the $T_{\rm dust}$–$\beta$ degeneracy present in far-infrared–only analyses. 
For typical dust temperatures of $\sim$18 K in M31, the {\it Herschel} bands do not fully cover the Rayleigh–Jeans limit, whereas the additional millimeter coverage provides improved leverage on the long-wavelength SED slope.
\item We derive an average $\beta \approx 1.8$ in the mapped region, lower than values reported in some previous {\it Herschel}-based studies. 
The enhanced millimeter flux relative to far-infrared extrapolations leads to a flatter fitted SED slope and correspondingly smaller $\beta$ values.
\item Although the 5 kpc ring and the 10 kpc ring exhibit similar dust temperatures, the 5 kpc ring shows a higher median $\beta$ value of 2.03 and a broader distribution compared to the 10 kpc ring with a median $\beta$ value of 1.71. 
Statistical tests indicate that the two samples are unlikely to be drawn from the same parent distribution, suggesting possible environmental variations in dust properties.
The elevated $\beta$ values in the inner 5 kpc may be related to variations in dust composition, possibly driven by enhanced starlight heating in this region.
\end{enumerate}

Overall, these findings highlight the importance of extended wavelength coverage for constraining dust emissivity in nearby galaxies and for probing environmental variations.
For dust temperatures typical of M31 ($\sim$18 K), far-infrared data alone do not fully sample the Rayleigh–Jeans regime, which might introduce systematic uncertainties in $\beta$ fitting. 
The inclusion of millimeter observations provides improved constraints on the long-wavelength SED slope, and similar limitations may affect studies of other galaxies that lack millimeter data.
A dedicated analysis of the complementary NIKA2 2 mm data will be presented in future work.

\begin{acknowledgements}
    This work is based on observations carried out under project number 066-25 with the IRAM 30m telescope. 
    IRAM is supported by INSU/CNRS (France), MPG (Germany) and IGN (Spain).
    S.J. is supported by NSFC grant nos. 12588202, 12041302, and 12603047.
    D. L. acknowledges support from the New Cornerstone foundation.
    S.J. thanks Angel Bongiovanni and Ioannis Myserlis for helpful discussions on the observational strategy and observational operations.
\end{acknowledgements}

\bibliographystyle{aa}
\bibliography{reference}

\begin{appendix}  

\section{CO(2–1) contamination in the 1.15 mm continuum}\label{co_contamination}

The CO(2–1) line falls within the bandpass of the NIKA2 1.15 mm observations and can therefore contribute to the measured continuum emission. 
To assess the level of CO(2–1) contamination, we quantified the fraction of line flux contributing to the continuum following \citet{Drabek2012MNRAS,Pantoni2026}, using a conversion factor $C$.

\begin{figure}
\vspace{0.01cm}
\includegraphics[width=8.7cm]{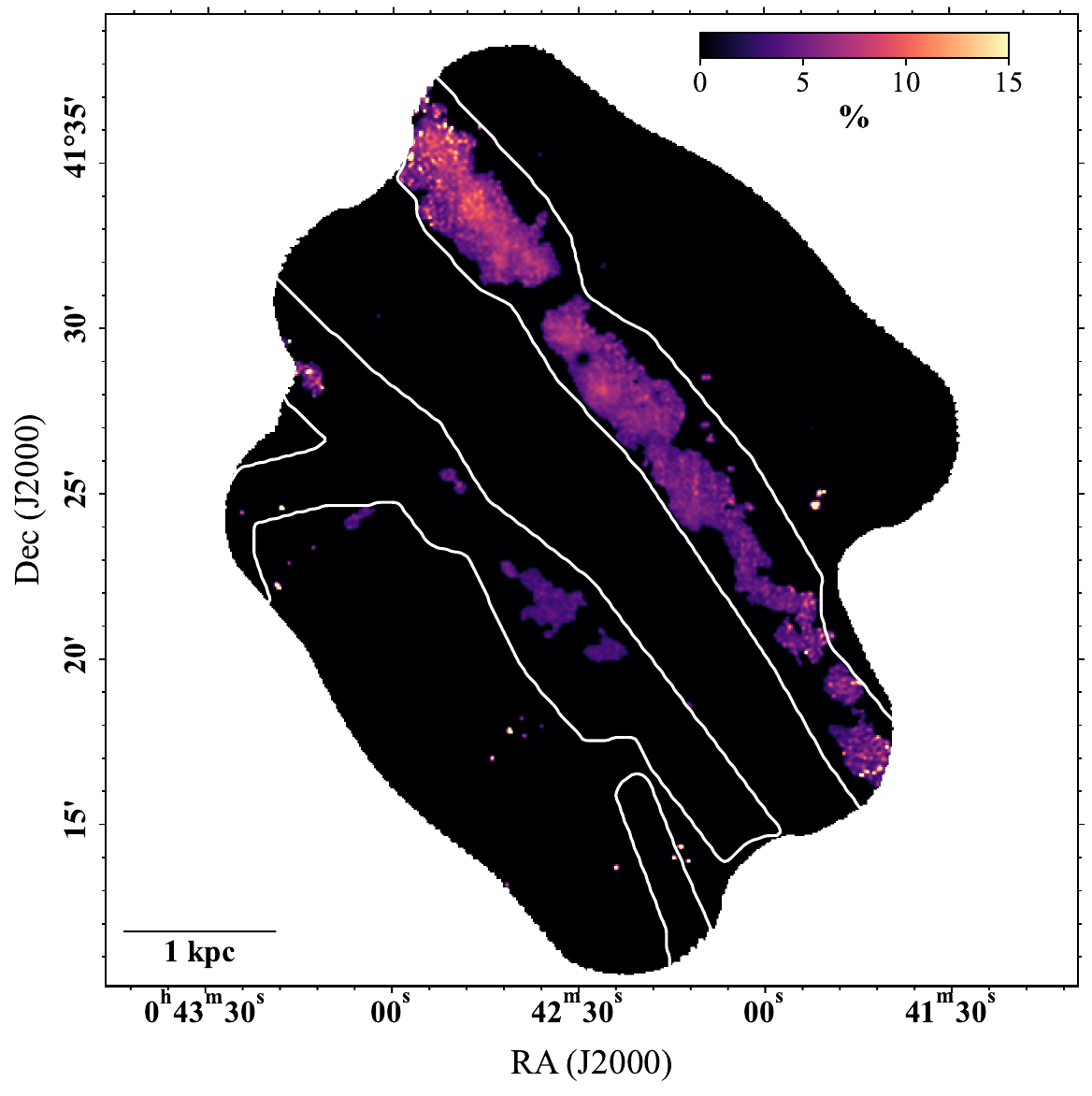}
\caption{
Percentage of flux in the NIKA2 1.15 mm map due to CO(2–1) contamination.
The white contours trace the spiral-arm mask from \citet{Jiao2025am31}.
}
\label{fig:percentage}
\end{figure}

However, no CO(2–1) map with angular resolution comparable to our NIKA2 observations is currently available for the field of view studied here.
We therefore adopted the predicted CO(3–2) integrated intensity map from \citet{SmithApJS} and converted it to a CO(2–1) intensity map by assuming a CO(3–2)/CO(2–1) line ratio of $\sim$0.3.
This value is based on previously reported CO(2–1)/(1–0) line ratio of $\sim$0.65 and CO(3–2)/(1–0) line ratio of $\sim$0.14–0.27 across the disk of M31 \citep{Nieten2006A&A,Li2020MNRAS}.
The resulting predicted CO(2–1) map has an angular resolution of $\sim$13$''$. 
At this resolution, the corresponding contamination conversion factor is $C = 0.05$ mJy beam$^{-1}$ per K km s$^{-1}$.
Using this map, we estimated that the typical CO(2–1) contamination flux in the ring region is $\sim$0.2 mJy beam$^{-1}$, which is lower than the noise level of our 1.15 mm map. 
We further calculated the percentage of flux in the NIKA2 1.15 mm map due to CO(2–1) contamination at 13$''$ resolution, as shown in Figure~\ref{fig:percentage}. 
For the majority of pixels (99.7$\%$), the contamination fraction is below 10$\%$, indicating that the CO(2–1) contribution to the continuum emission is generally small in the target region.
To assess the impact on the SED fitting, we repeated the analysis using a NIKA2 1.15 mm map with the predicted CO(2–1) contamination subtracted. 
The resulting dust properties show only minor differences.
In particular, for $\Sigma_{\rm dust}$, the difference is smaller than 10$\%$ for 99.3$\%$ of pixels compared to the results obtained without subtracting the CO(2–1) contribution.

Nevertheless, we note that the present estimate relies on several assumptions, including the use of a predicted CO(3–2) map and an assumed line ratio to infer the CO(2–1) intensity. 
These approximations introduce significant uncertainties. 
Direct CO(2–1) observations at comparable angular resolution would therefore be important to more accurately quantify and subtract the CO(2–1) contamination from the continuum map.
 
\section{Comparison with the full-band SED fitting}
\label{appendix:fullband}
 
In the main analysis, we adopt the results obtained from fitting the 250--1150~$\mu$m data.
Excluding the 70--160~$\mu$m bands reduces the sensitivity of the fitted parameters to warmer dust components.
For comparison, we also performed the iterative SED fitting using the full 70--1150~$\mu$m wavelength coverage, following the same procedure described in Section~\ref{section:obs_sed}.

\begin{figure*}[ht!]
\vspace{-0.3cm}
\hspace{-0.1cm}
\begin{center}
\includegraphics[width=16.cm,]{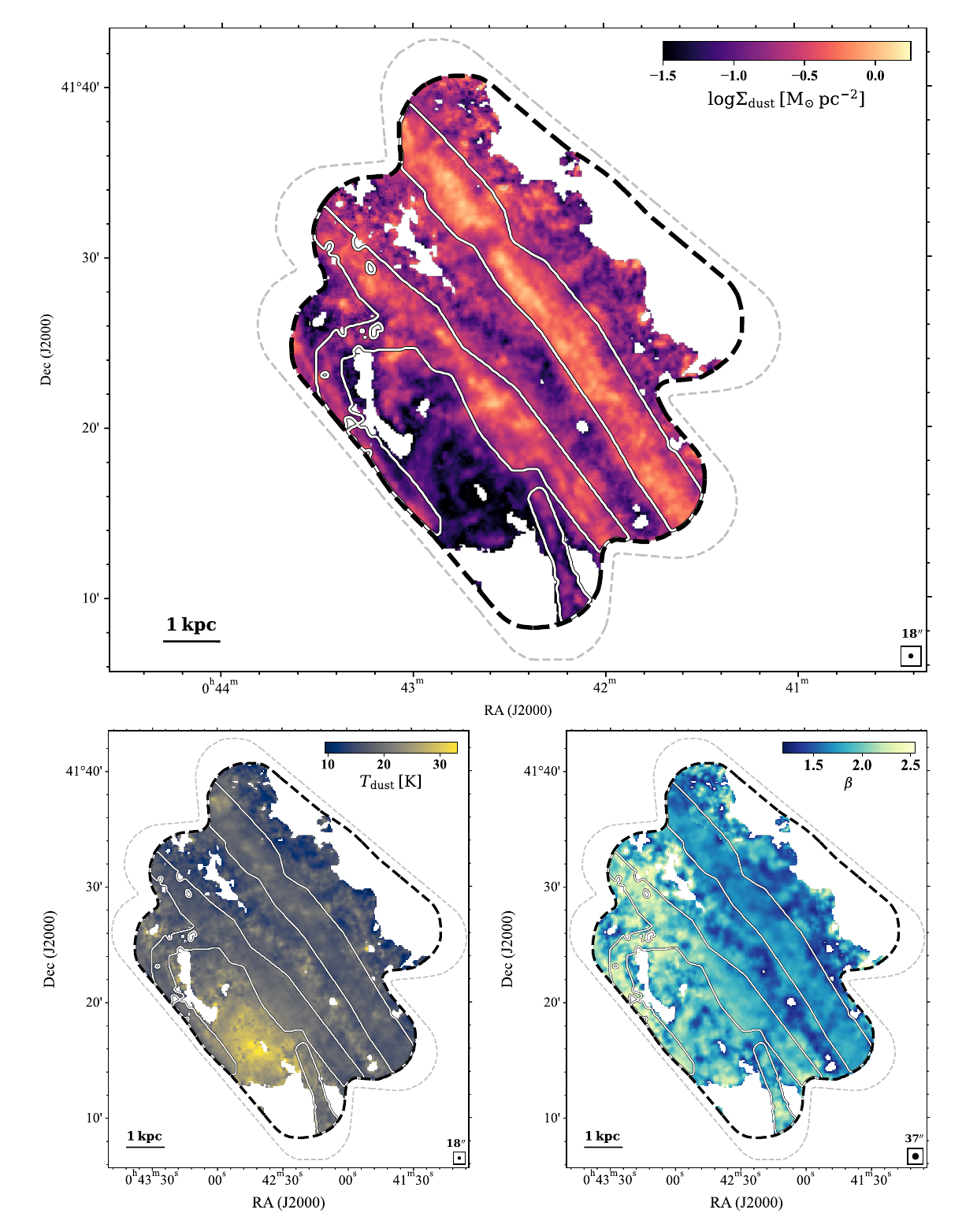}
\end{center}
\vspace{-0.3cm}
\caption{
The distribution of dust surface density (18$''$ angular resolution), temperature (18$''$ angular resolution), and $\beta$ (37$''$ angular resolution) obtained from the full-band iterative SED fits.
The gray dashed line delineates the field of view of the new NIKA2 observations, while the black dashed contour corresponds to the 2 mJy beam$^{-1}$ sensitivity limit of the NIKA2 1.15 mm data.
The white contours trace the spiral-arm mask from \citet{Jiao2025am31}.}
\label{fig:sed_maps_fullband}
\end{figure*}

Figure~\ref{fig:sed_maps_fullband} presents the dust surface density, temperature, and emissivity-index maps obtained from the full-band fits.
The overall morphology is broadly similar to that recovered from the 250--1150~$\mu$m fits shown in Figure~\ref{fig:sed_maps}.
In both cases, the enhanced dust surface density traces the 5 and 10 kpc rings, while the dust temperature generally decreases from the inner region toward the disk.
The full-band fit also recovers coherent spatial variations in $\beta$, although the absolute fitted values and their dispersion differ to some extent from those obtained from the long-wavelength-only fit.

\begin{figure}
\vspace{0.1cm}
\includegraphics[width=9.cm]{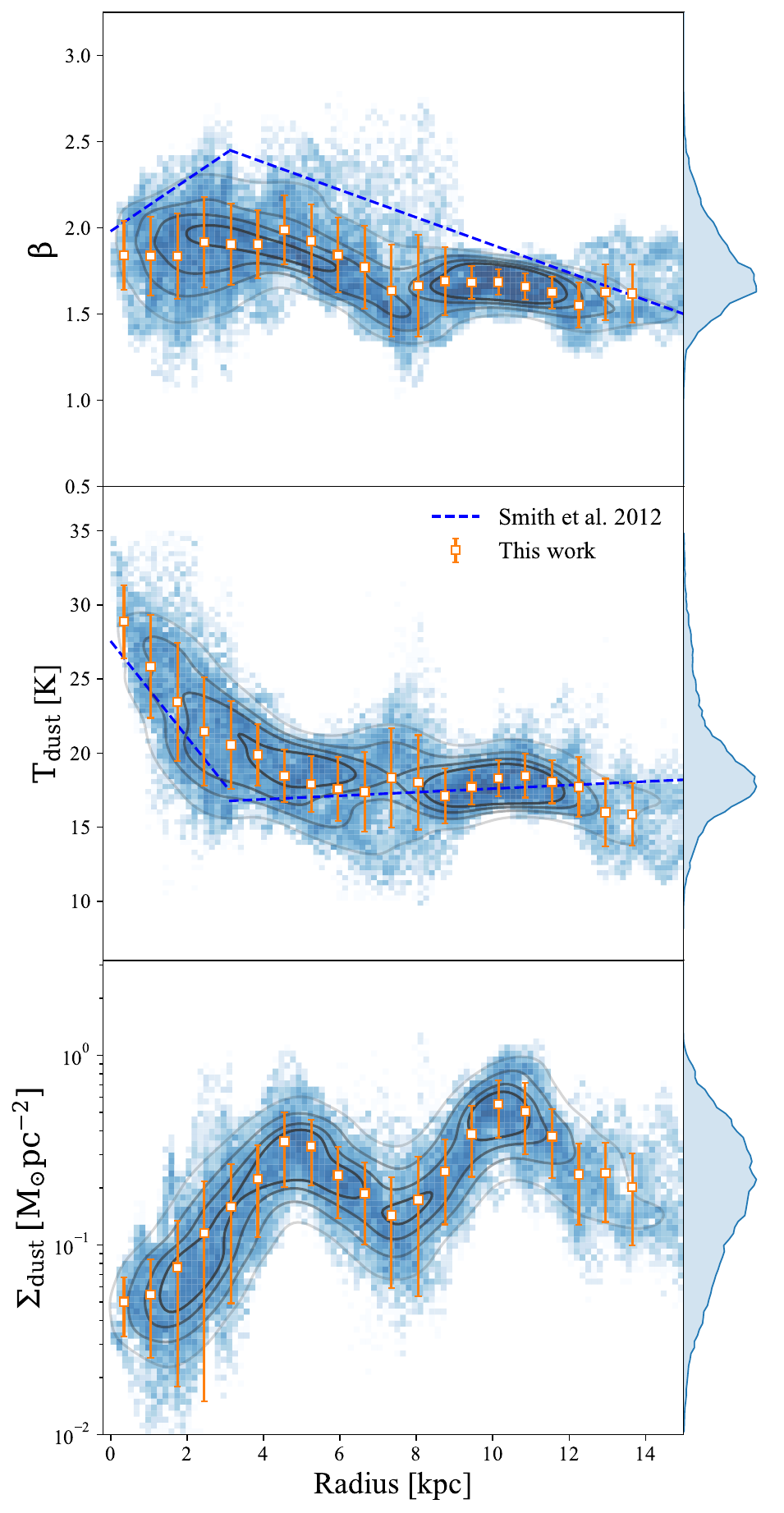}
\caption{
Results from the full-band iterative SED fits for each pixel plotted vs. radius.
The blue points show the fitted value for each pixel, while the yellow squares represent averages in 700 pc bins.
The corresponding yellow error bars represent the standard deviation within each bin.
The dashed blue lines represent the fitted relationships from the HELGA II \citep{Smith2012ApJ}.
The histograms on the right side show the dust surface density, dust temperature, and $\beta$ distribution of the map region shown in Figure \ref{fig:sed_maps}.
}
\label{fig:fullband_radial}
\end{figure}
 
Figure~\ref{fig:fullband_radial} shows the corresponding radial distributions of $\Sigma_{\rm dust}$, $T_{\rm dust}$, and $\beta$ derived from the full-band SED fitting.
The radial behavior of the fitted parameters is qualitatively consistent with that obtained from the 250--1150~$\mu$m analysis.
In particular, both wavelength configurations recover the surface-density enhancements associated with the 5 and 10 kpc rings and a general decline in dust temperature from the inner region toward the disk.
They also show systematically higher effective $\beta$ values around the 5 kpc ring than around the 10 kpc ring.
For the full-band fit, the median $\beta$ values in the 5 and 10 kpc
rings are 1.92 and 1.67, respectively, compared with 2.03 and 1.71 from the 250--1150~$\mu$m fit.

\end{appendix}

\end{document}